\documentclass[floats,floatfix,showpacs,preprintnumbers,amssymb,prd,superscriptaddress,nofootinbib,reprint]{revtex4-2}

\usepackage{graphicx}
\usepackage{dcolumn}
\usepackage{bm}
\usepackage{amsmath}
\usepackage{amssymb}
\usepackage{physics}
\usepackage{tensor}
\usepackage[colorlinks=true]{hyperref}
\usepackage{mathtools}

\usepackage{xcolor}
\usepackage[normalem]{ulem}

\usepackage{booktabs}
\usepackage{tikz}
\usepackage{pgfplots}
\pgfplotsset{compat=1.14}

\newcommand{\UVA}{Department of Physics, University of Virginia, P.O.~Box 400714, Charlottesville, VA 22904-4714, USA}

\newcommand{\MSU}{eXtreme Gravity Institute, Department of Physics,
Montana State University, Bozeman, MT 59717, USA}

\begin{document}


\title{Anisotropic Fluid Model for Massive Boson Stars: \\ Spherically Symmetric Configurations}

\author{Ethan McKeever}
 \affiliation{\MSU}

 \author{Victor Guedes}
 \affiliation{\UVA}

 \author{Shu Yan Lau}
 \affiliation{\MSU}

\author{Kent Yagi}
 \affiliation{\UVA}

\begin{abstract}

Boson stars are among the best-studied exotic compact objects and constitute one of the few black-hole mimickers for which both the formation mechanism and dynamical behavior can be investigated from first principles. Constructing boson star solutions is generally more challenging than constructing neutron star solutions, often requiring either extremely accurate initial guesses for the oscillation frequency or the use of computationally intensive relaxation methods to solve the field equations.
In the strong-coupling limit, massive boson stars can be described effectively as perfect-fluid configurations, enabling their construction using methods analogous to those developed for neutron stars. In this paper, we relax the strong-coupling assumption and derive a phenomenological anisotropic-fluid model for spherically symmetric massive boson stars that remains applicable over a broad range of model parameters. We provide ready-to-use fits for both the anisotropy profile and the effective equation of state in terms of the central scalar-field amplitude and the ratio of the self-interaction coupling constant to the squared scalar-field mass.
To validate the model, we compute the masses and radii of boson stars using a standard neutron-star construction approach and compare the results against full boson-star calculations. We find excellent agreement between the two methods, while the new approach reduces the computational cost by a factor of $\sim10$. Our phenomenological fluid framework offers an efficient and accurate method for constructing spherically symmetric massive boson stars and computing their observable properties. We provide a ready-to-use Python implementation of the Tolman-Oppenheimer-Volkoff solver, allowing users to construct boson star configurations within the proposed framework.
\end{abstract}

\maketitle

\section{\label{sec:level1} Introduction}

Gravitational waves were first detected from a binary black hole merger on September 14, 2015, by the Advanced LIGO observatories \cite{PhysRevLett.116.061102,2015}. Less than two years later, the Advanced LIGO and Advanced Virgo detectors~\cite{Acernese_2014} observed the first binary neutron star inspiral~\cite{PhysRevLett.119.161101}. Since then, the global detector network has expanded to include KAGRA~\cite{Aso_2013}, completed its fourth observing run, and made gravitational-wave detections routine~\cite{LIGOScientific:2018mvr,LIGOScientific:2020ibl,KAGRA:2021vkt}. The latest catalog, GWTC-5.0, contains 390 confirmed gravitational-wave events~\cite{LIGOScientific:2026wfs}.

Although all observed compact binary mergers are consistent with black holes and neutron stars, the existence of exotic compact objects has not been ruled out~\cite{Krishnendu:2025rud}. Proposed alternatives include boson stars~\cite{Schunck_2003,Liebling_2023}, quark stars~\cite{10.1143/PTP.44.291,Weber_2012}, gravastars~\cite{Mazur_2023,Mazur_2004}, and axion stars~\cite{Iwazaki_1999,Eby_2017}; see Ref.~\cite{Cardoso:2019rvt} for a comprehensive review. Among these candidates, boson stars are particularly compelling as the discovery of the Higgs boson~\cite{Aad_2012,201230} established the existence of fundamental scalar fields in nature, while the gravitational cooling mechanism provides a generic formation channel for self-gravitating scalar configurations~\cite{PhysRevLett.72.2516}.

In this work, we focus on massive boson stars~\cite{PhysRevLett.57.2485}, described by a scalar field with a quadratic mass term and a quartic self-interaction potential. Their stability against radial perturbations has been investigated by Jetzer and Gleiser~\cite{JETZER1989411,PhysRevD.38.2376}, while their tidal deformability~\cite{Sennett_2017,Cardoso_2017}, binary merger dynamics and gravitational-wave emission~\cite{PhysRevD.77.044036,PhysRevD.96.104058}, and quasinormal-mode spectrum~\cite{PhysRevD.50.6235,PhysRevD.58.104004,Macedo:2013jja} have all been studied extensively. In addition, Kling and Rajaraman~\cite{PhysRevD.97.063012} derived a semi-analytic solution that reproduces fully numerical calculations with high accuracy.

Despite this progress, constructing equilibrium boson star solutions remains significantly more involved than constructing neutron star models. The Einstein--Klein--Gordon equations must be solved as a boundary value problem. The most widely used approach is the shooting method~\cite{Press:2007ipz}, although it often requires an extremely accurate initial guess for the scalar-field oscillation frequency~\cite{Macedo:2013jja}. Alternatively, one may employ a relaxation method, which is generally more robust but also more computationally expensive~\cite{Press:2007ipz}.

A notable simplification occurs in the strong-coupling (Thomas--Fermi) limit, where the scalar self-interaction dominates over the mass term. In this regime, Colpi \textit{et al.}~\cite{PhysRevLett.57.2485} showed that the scalar-field stress-energy tensor can be written in the form of a perfect fluid. Consequently, equilibrium configurations can be obtained by solving an effective Tolman--Oppenheimer--Volkoff (TOV) system, analogous to neutron stars~\cite{Chavanis:2011cz}. This effective-fluid description has enabled studies of rotating boson stars~\cite{Ryan:1996nk,Vaglio:2022flq}, universal relations~\cite{Maselli:2017vfi}, and oscillation modes~\cite{Flores_2019}.

The objective of this paper is to extend the applicability of this fluid description beyond the strong-coupling regime. For generic massive boson stars, the scalar-field stress-energy tensor is known to take the form of an anisotropic fluid~\cite{Macedo:2013jja}. Motivated by this observation, we develop a phenomenological anisotropic-fluid model by constructing ready-to-use expressions for the pressure anisotropy profile and the effective equation of state as functions of the central scalar-field amplitude and the ratio of the self-coupling constant to the squared boson mass. Like the perfect-fluid model in the Thomas--Fermi limit, our formulation can be readily incorporated into existing TOV solvers (e.g.,~\cite{2024arXiv240411346K}). We also provide a public implementation of our solver~\cite{OnlineLink}, enabling efficient construction of massive boson star models over a broad range of coupling strengths.

To assess the accuracy of the proposed model, we reconstruct spherically symmetric boson stars and compare the results with full numerical solutions obtained using the relaxation method. Across the parameter space considered, our phenomenological model reproduces the stellar mass and radius with maximum errors of approximately $4\%$ and $7\%$, respectively. Here, the radius in the full numerical calculation is defined as the radius enclosing $99\%$ of the total mass. Using the radius predicted by our phenomenological model, the enclosed mass in the full solution ranges from 98\% to 99\% of the total mass, depending on the boson star parameters.

The remainder of this paper is organized as follows. In Sec.~\ref{Boson Stars}, we review the massive boson star model and describe the numerical methods used to solve the field equations. In Sec.~\ref{Anisotropic Model}, we present analytic fits for the pressure anisotropy over a broad range of boson self-couplings and central scalar-field amplitudes. In Sec.~\ref{Applications}, we derive approximate expressions for the stellar mass and radius using our phenomenological model and compare them with the full numerical solutions to demonstrate its accuracy.

We use the metric signature $(-, +, +, +)$ and use Planck natural units $\hbar = c = G = 1$.
We sometimes restore factors of $\hbar$, $c$, and the Planck mass $m_\mathrm{Planck} = \sqrt{\hbar c /G}$ to improve clarity. We use Greek indices $(\alpha, \beta, \cdots)$ to represent spacetime indices and we define the Riemann curvature tensor as $\nabla_{\beta} \nabla_{\alpha} a_{\mu} - \nabla_{\alpha} \nabla_{\beta} a_{\mu} = {R}\indices{^{\nu}_{\mu}_{\alpha}_{\beta}} a_{\nu}$, where $\nabla_{\alpha}$ is the covariant derivative and $a_{\mu}$ is a generic covector.

\section{Boson Stars}
\label{Boson Stars}

In this section, we review the basics and how to construct spherically symmetric solutions for massive boson stars. 

\subsection{Action and Field Equations}

Boson stars are stellar objects composed solely of scalar fields and are held up by pressure associated with the Heisenberg uncertainty principle. They were first theoretically demonstrated by Ruffini and Bonazzola~\cite{PhysRev.187.1767}. They are described by the Einstein-Klein-Gordon action~\cite{Macedo:2013jja}
\begin{equation}
    S = \int d^4x\sqrt{-g}\left[\frac{R}{16\pi G}-\nabla ^{\alpha}\Phi\nabla_{\alpha}\Phi ^* - V(|\Phi|^2)\right],
\end{equation}
where $g$ is the metric determinant, $R$ is the Ricci scalar, $\Phi$ is the complex scalar field, $^*$ denotes complex conjugation, and $V(|\Phi|^2)$ is a given potential. The specific form of this potential describes the model of boson stars. In this paper, we focus on ``massive" boson stars~\cite{PhysRevLett.57.2485} which have the potential
\begin{equation}
    V(|\Phi|^2) = \mu ^2 |\Phi| ^2 + \frac{\lambda}{2}|\Phi| ^4,\label{potential}
\end{equation}
 where $\mu$ is the mass of the boson and $\lambda$ is the self-interaction coupling. This potential is repulsive for $\lambda~\geq~0$. 

The Einstein and Klein-Gordon equations for this action are given as
\begin{eqnarray}
\label{eq:einstein}
    R_{\alpha \beta}-\frac{1}{2}g_{\alpha \beta}R &=& 8\pi T_{\alpha \beta}, \\
    \nabla^{\alpha}\nabla_{\alpha}\Phi &=& \frac{dV}{d|\Phi|^2}\Phi, \label{eq:klein_gordon}
\end{eqnarray}
with the stress-energy tensor for the scalar field given by~\cite{Sennett_2017}
\begin{align}
\label{eq:Tmunu}
    T_{\alpha \beta} =  2\nabla_{(\alpha}\Phi^*\nabla_{\beta)}\Phi 
    - g_{\alpha \beta}(\nabla^{\gamma}\Phi^*\nabla_{\gamma}\Phi + V(|\Phi|^2)).
\end{align}
Solving Eqs.~\eqref{eq:einstein}--\eqref{eq:Tmunu} grants us the dynamics of the metric and the scalar field.

\subsection{Spherically Symmetric Solutions}

To construct static and spherically symmetric boson star solutions, we solve the above field equations under some ansatz. For the metric, we assume the following spherically symmetric form:
\begin{equation}
    ds^2 = -e^{v(r)}dt^2+e^{u(r)}dr^2+r^2(d\theta^2+\sin^2\theta\, d\varphi^2).
\end{equation}
We then assume the following form for the scalar field: 
\begin{equation}
    \Phi(t,r) = \phi_0(r)e^{-i\omega t},
\end{equation}
where $\omega$ is the oscillation frequency of the boson field. 

For the spherically symmetric case, the stress-energy tensor for the scalar field in Eq.~\eqref{eq:Tmunu} can be mapped to that of an anisotropic fluid. The effective energy density $\rho$, radial pressure $p_r$, and tangential pressure $p_\perp$ are given as~\cite{Macedo:2013jja,Sennett_2017}
\begin{align}
     \rho \equiv & -{T}\indices{^{t}_{t}} = \omega^2 e^{-v}\phi_0^2+e^{-u}(\phi'_0)^2+V_0, \\
p_{r} \equiv & \, {T}\indices{^r_r} = \omega^2e^{-v}\phi_0^2+e^{-u}(\phi'_0)^2-V_0, \\
p_{\perp} \equiv & \, {T}\indices{^{\theta}_{\theta}} = \omega^2e^{-v}\phi_0^2-e^{-u}(\phi'_0)^2-V_0,
\end{align}
where $V_0 = V(\phi_0)$ and a prime denotes a radial derivative. The field equations then reduce to~\cite{Sennett_2017}
\begin{eqnarray}
\label{eq:field1}
    e^{-u}\left(-\frac{u'}{r}+\frac{1}{r^2}\right)-\frac{1}{r^2} &=& -8\pi \rho,\\
\label{eq:field2}
    e^{-u}\left(\frac{v'}{r}+\frac{1}{r^2}\right)-\frac{1}{r^2} &=& 8\pi p_{r}, \\
\label{eq:field3}
    \phi_0''+\left(\frac{2}{r}+\frac{v'-u'}{2}\right)\phi_0'&=&e^u\left(U_0-\omega^2e^{-v}\right)\phi_0,
\end{eqnarray}
where $U_0 = U(\phi_0)$ with $U =  dV/d|\Phi|^2$ and a prime denotes a radial derivative.
 We solve the above field equations using the relaxation method (see~\cite{Sennett_2017,Press:2007ipz} for details on the method) under the following boundary conditions:
 \begin{align}
    \lim_{r \to 0} m(r) &\sim r^3, & \quad \lim_{r \to r_{\rm max}} m(r) &\sim M, \\
    \lim_{r \to 0} v(r) &\sim v_c, & \quad \lim_{r \to r_{\rm max}} v(r) &\sim 0, \\
    \lim_{r \to 0} \phi_0(r) &\sim \phi_{0c}, & \quad \lim_{r \to r_{\rm max}} \phi_0(r) &\sim \frac{1}{r} e^{-r\sqrt{\mu^2 - \omega^2}}.
\end{align}
where a subscript $c$ denotes a value at stellar center and, technically, $r_{\rm max}\rightarrow\infty$, since the scalar field extends to infinity. We rescale the radial coordinate $r$, the self-coupling constant $\lambda$, and the scalar field function $\phi_0$ in terms of the boson mass $\mu$ as follows~\cite{Macedo:2013jja,Sennett_2017}:
\begin{align}
\label{rscale}
    r \rightarrow & \frac{m_\mathrm{Planck}^2\tilde{r}}{\mu}, \\
   \label{lambdascale}
   \lambda \rightarrow & \frac{8\pi \mu ^2 \tilde{\lambda}}{m^2_\mathrm{Planck}}, \\
 \label{phiscale}
\phi_0 \rightarrow &\frac{m_\mathrm{Planck}\tilde{\phi_0}\tilde{r}}{(8\pi)^{1/2}}.
\end{align}
We assume $\mu=1$, without any loss of generality. We solve Eqs.~\eqref{eq:field1}$-$\eqref{eq:field3} up to ${\tilde{r}}_{\rm max}=75$, which we found to be a good approximation that does not affect the stellar radius $R$, defined as the radius of a sphere inside which 99\% of the total mass $M$ is enclosed.

\begin{figure}
\includegraphics[width=8.6cm]{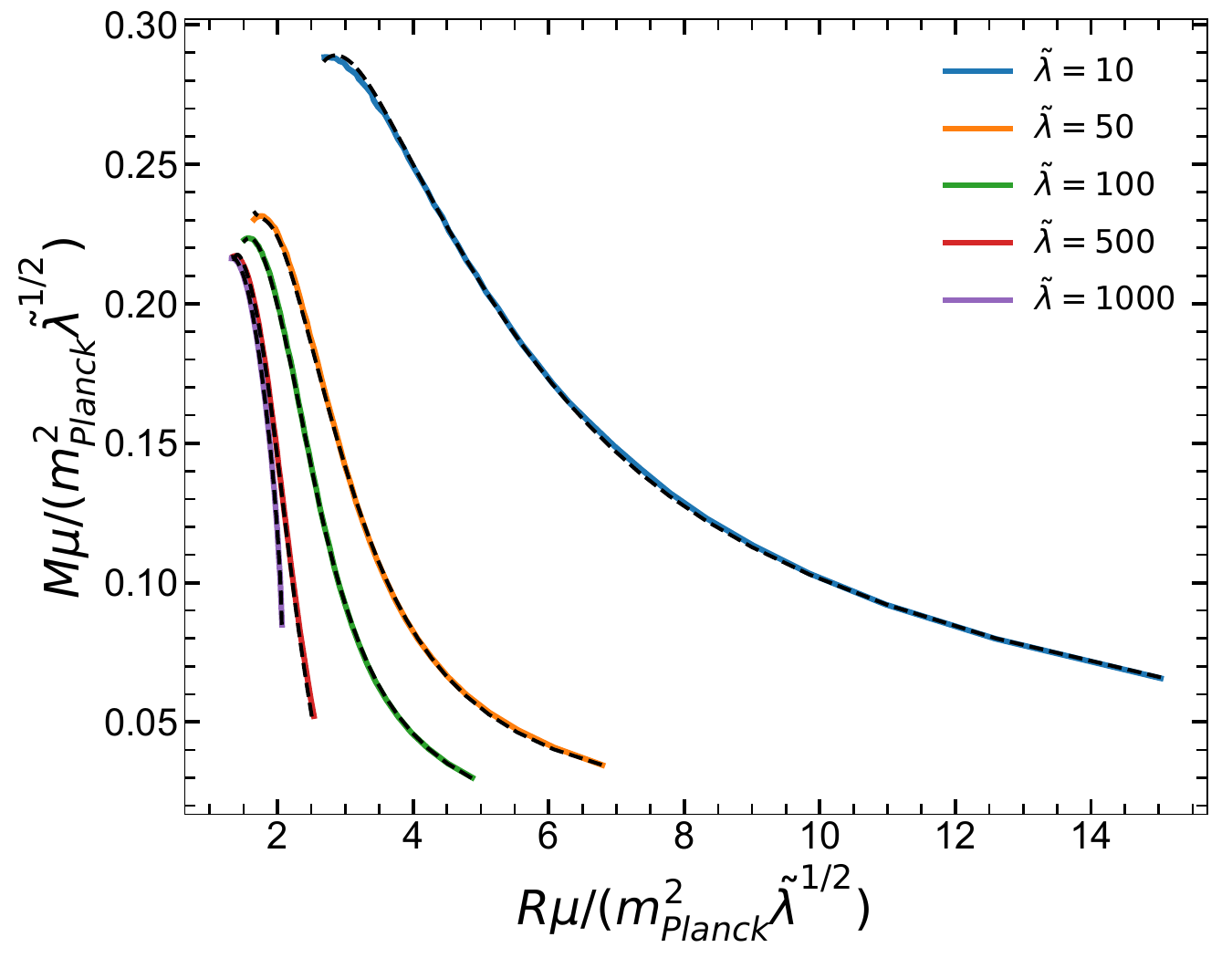}
\caption{\label{fig:MassRadius} 
Mass-radius relations for massive boson stars described by various self-interaction couplings (see Eqs.~\eqref{potential} and~\eqref{lambdascale}). Solid colored lines show values obtained using the relaxation method while the black dashed lines overlaying each solid line show values obtained using our phenomenological model.
}
\end{figure}

The mass-radius relations for boson stars are shown in Fig.~\ref{fig:MassRadius} as solid colored curves. We choose $\tilde \lambda$ in the range 0--1000. For each $\tilde \lambda$, we vary the central value of the scalar field, $\tilde \phi_{0c}$, from 0.01 to the value that grants the maximum-mass stellar configuration.
For example, in the strong coupling limit of $\lambda \gg \mu^2/m_\mathrm{Planck}^2$ or $\tilde{\lambda} \gg 1$, spherically symmetric massive boson stars have a maximum mass \cite{PhysRevLett.57.2485,Sennett_2017} 
\begin{align}
    M_{\max} &\approx 0.044 \, \sqrt{\lambda} \frac{m^3_\mathrm{Planck}}{\mu^2} \nonumber \\
 &\approx 1 \, {M}_\odot \, \sqrt{\lambda}\left(\frac{0.3 \, \mathrm{GeV}}{\mu}\right)^2\,.
\end{align}
Stable configurations of boson stars have a maximum compactness $C_{\max} \approx 0.158$ in the strong coupling limit (roughly as compact as neutron stars) and $C_{\max} \approx 0.08$ for $\lambda = 0$~\cite{Guzm_n_2006, Amaro_Seoane_2010}.

\section{Phenomenological Anisotropic Fluid Models for Boson Stars}
\label{Anisotropic Model}

In the strong coupling limit, massive boson stars are known to follow an isotropic perfect fluid form with the effective equation of state given by (putting back the speed of light $c$ for convenience)~\cite{PhysRevLett.57.2485}
\begin{equation}
    p = \frac{c^4}{36K}\left[\left(1+\frac{12K}{c^2}\rho\right)^{\frac{1}{2}}-1\right]^2, \quad K \equiv \frac{\lambda\hbar^3}{8 \mu^4c}.
\end{equation}
The goal of this paper is to find a phenomenological anisotropic fluid model for boson stars that is applicable beyond the strong coupling limit. The difference  between the radial and tangential pressure for boson stars is given by
\begin{equation}
    \sigma \equiv p_{r}-p_{\perp} = 2e^{-u}(\phi'_0)^2. \label{eq_aniso}
\end{equation}

\subsection{Anisotropy Profile}

Various phenomenological anisotropic fluid models have been proposed for neutron stars. For example, 
Herrera and Barreto (HB)~\cite{Herrera_2013,Setiawan:2019ojj} proposed a model (initially studied by Cosenza et al.~\cite{10.1063/1.524742}) 
 \begin{equation} \label{eq:sigma_HB}
     \sigma = \frac{\beta_\mathrm{HB}}{2(\beta_\mathrm{HB} +1)}\tilde r \frac{dp_r}{d\tilde r},
\end{equation}
where $\beta_\mathrm{HB}$ is a parameter characterizing anisotropy. The model reduces to the isotropic one in the limit $\beta_\mathrm{HB} \to 0$.
In this paper, we modify the above HB model in Eq.~\eqref{eq:sigma_HB} such that it fits better for massive boson stars\footnote{In addition to the model proposed by Herrera and Barreto (HB)~\cite{Herrera_2013}, we studied models proposed by Horvat (H) et al.~\cite{Horvat:2010xf} and Bowers and Liang (BL)~\cite{Bowers:1974tgi}. We found that HB-based and H-based models were able to fit our numerical results more accurately than the BL-based model. The HB and H models were more similar, with the H model being more accurate for $\tilde{\lambda} < 50$, while the HB model was more accurate for $\tilde{\lambda} > 50$. We chose to use the HB model for this work as the highest error from the HB model across our fitting regime was lower than that for the H model.}.

\begin{figure}
\includegraphics[width=8.6cm]{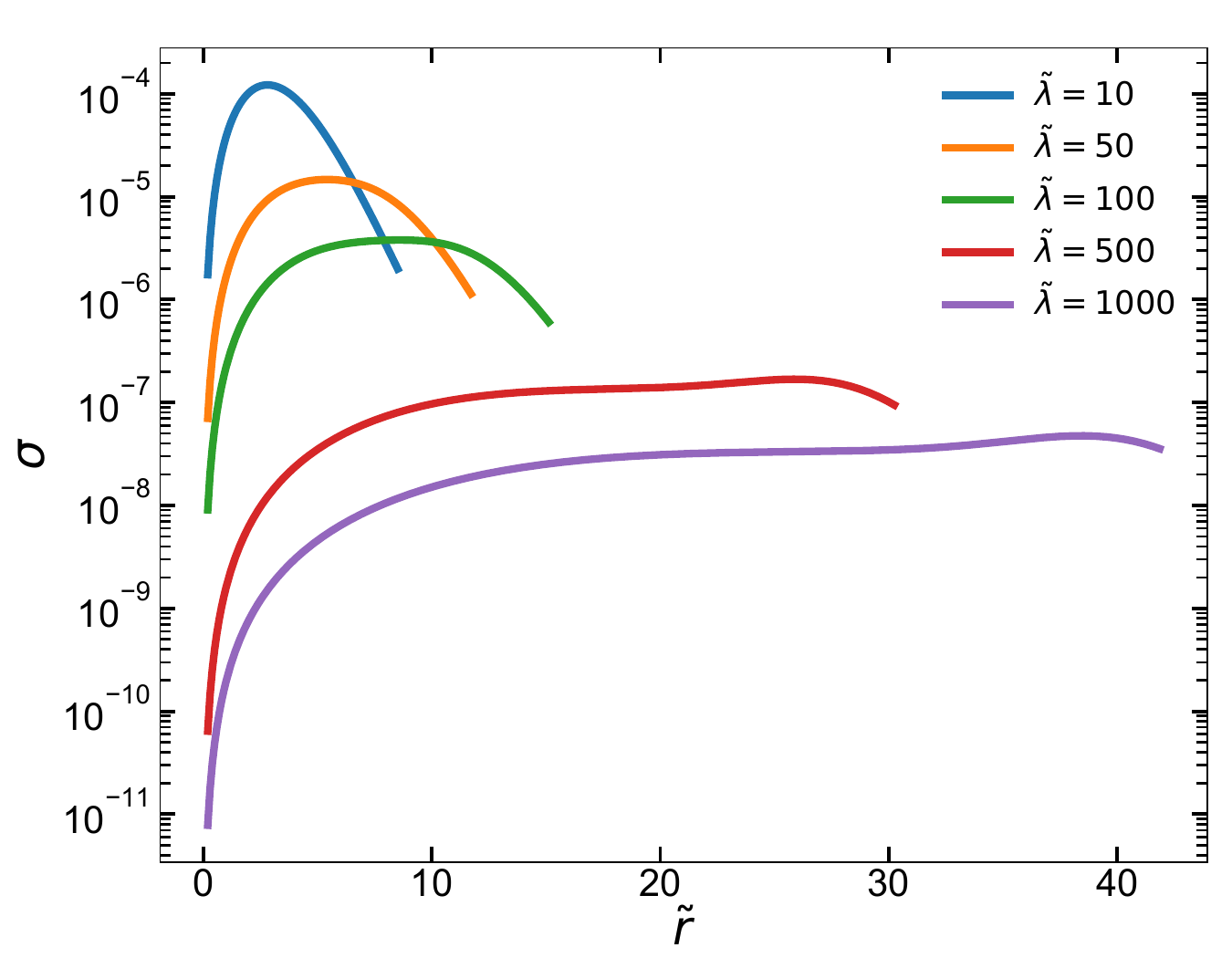}
\caption{\label{comparisonplot}Radial profiles of the effective pressure anisotropy (see Eq.~\eqref{eq_aniso}) for five models of boson stars near the maximum mass for the corresponding value of $\tilde{\lambda}$. These curves map to central scalar field values ${\tilde{\phi}}_{0c} = (0.23, 0.14, 0.10,0.044, 0.031)$ ordered in increasing $\tilde \lambda$.
}
\end{figure}

\begin{figure}
     \centering
         \includegraphics[width=0.4\textwidth]{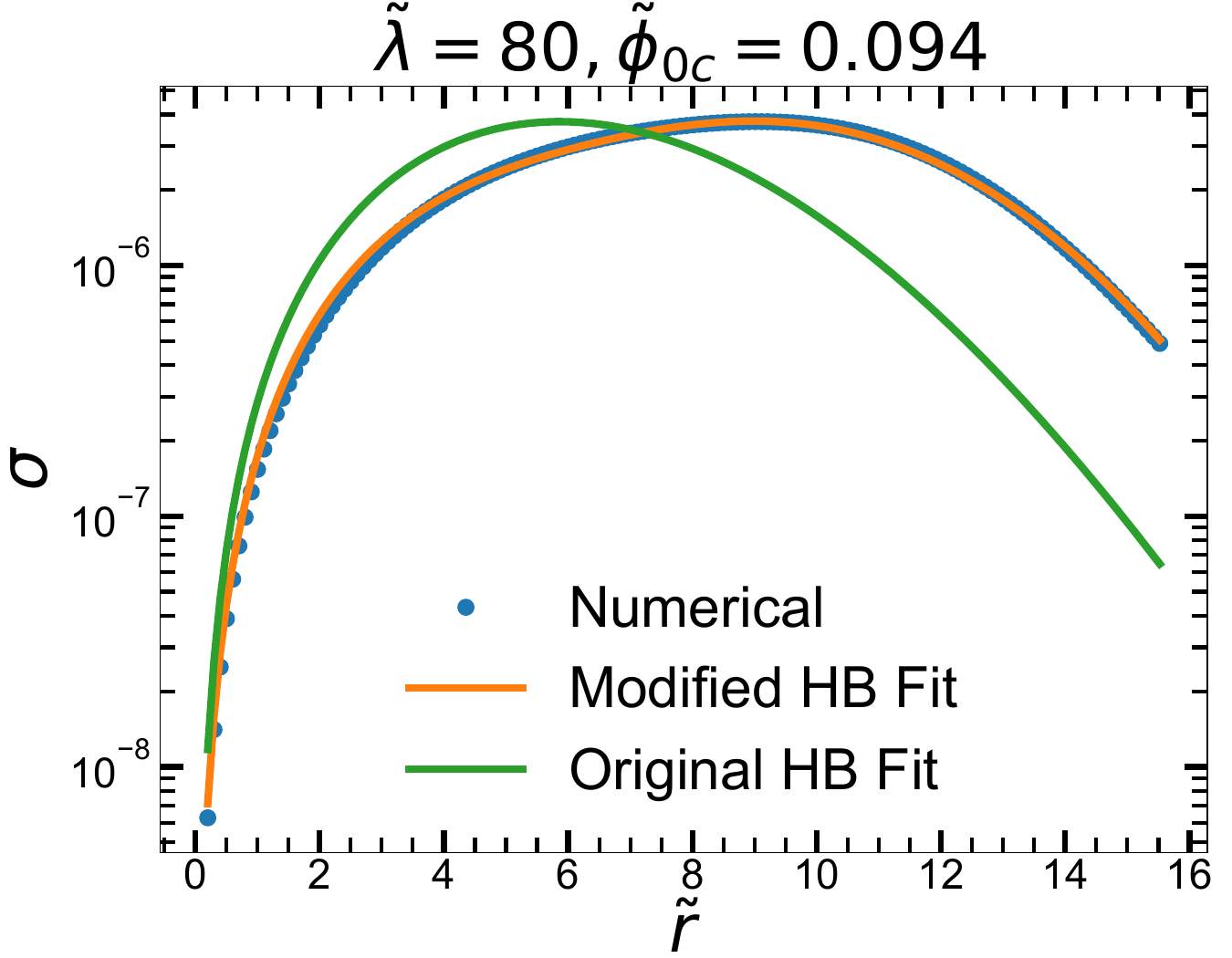}
         
        \caption{
Effective pressure anisotropy profile for boson stars obtained from the relaxation method (dots). We also present fits using the original (Eq.~\eqref{eq:sigma_HB}) and modified (Eqs.~\eqref{eq:sigma_ans} and~\eqref{eq:sigma_fit}) HB model.        
      }
    \label{fig:anisotropy}
\end{figure}

Anisotropy profiles of selected boson stars with different $\tilde \lambda$ are shown in Fig.~\ref{comparisonplot}. As expected, the anisotropy decreases as $\tilde \lambda$ increases (approaching the strong coupling limit). Due to the difference in their behavior (the peak of the profile shifting to a larger $\tilde r$ and small
fluctuations in the profiles appearing as we increase $\tilde \lambda$), we were unable to find a single function for $\sigma$ that works for a wide range of $\tilde \lambda$.  Hence, we divide the range of $\tilde \lambda$ into 3 regimes: (i) $0 \leq \tilde \lambda \leq 100$, (ii) $100 \leq \tilde \lambda \leq 500$, and (iii) $\tilde \lambda \geq 500$.  Within each regime, we modify the HB anisotropy model by introducing 3 or 4 anisotropy constants as follows:
\begin{equation}
\label{eq:sigma_ans}
    \sigma (\tilde{r}) = \frac{f(\tilde r)}{2(\beta_0 +1)} \frac{dp_r}{d\tilde r},
\end{equation}
with 
\begin{widetext}
\begin{equation}
\label{eq:sigma_fit}
     f(\tilde r) = \begin{dcases}
\tilde{r}(\beta_0 + \beta_1 \tilde{r}^{n} + \beta_2 \tilde{r}^{c_1  n})  & (0 \leq \tilde \lambda \leq 100), \\
\tilde{r}(\beta_0 + \beta_1 \tilde{r}^{n})  & (100 \leq \tilde \lambda \leq 500), \\
\beta_0 \tilde{r}^{c_2}+\beta_1 \tilde{r}^{n} + (\beta_2+c_3) [\ln(\tilde{r}+1)]^{c_4} & (\tilde \lambda \geq 500).
     \end{dcases}
      \end{equation}
\end{widetext}
Here\footnote{The shift in $\beta_2$ by $c_3$ for the high $\tilde \lambda$ range is introduced to force $\beta_2$ to be positive.}, $c_1 = 1.2$, $c_2=1.05$, $c_3=-6\times10^{-7}$, $c_4=11$, while the anisotropy constants $(\beta_0, \beta_1, \beta_2, n)$ in each region are fitted with the numerical profile for each combination of $(\tilde \lambda, \tilde \phi_{0c})$. 
Figure~\ref{fig:anisotropy} presents an example of the anisotropy profile inside the star and the corresponding fit. Note that, unlike the original HB model, the modified HB model above can accurately fit the numerical data. 

\begin{figure*}
     \centering
         \includegraphics[width=0.32\textwidth]{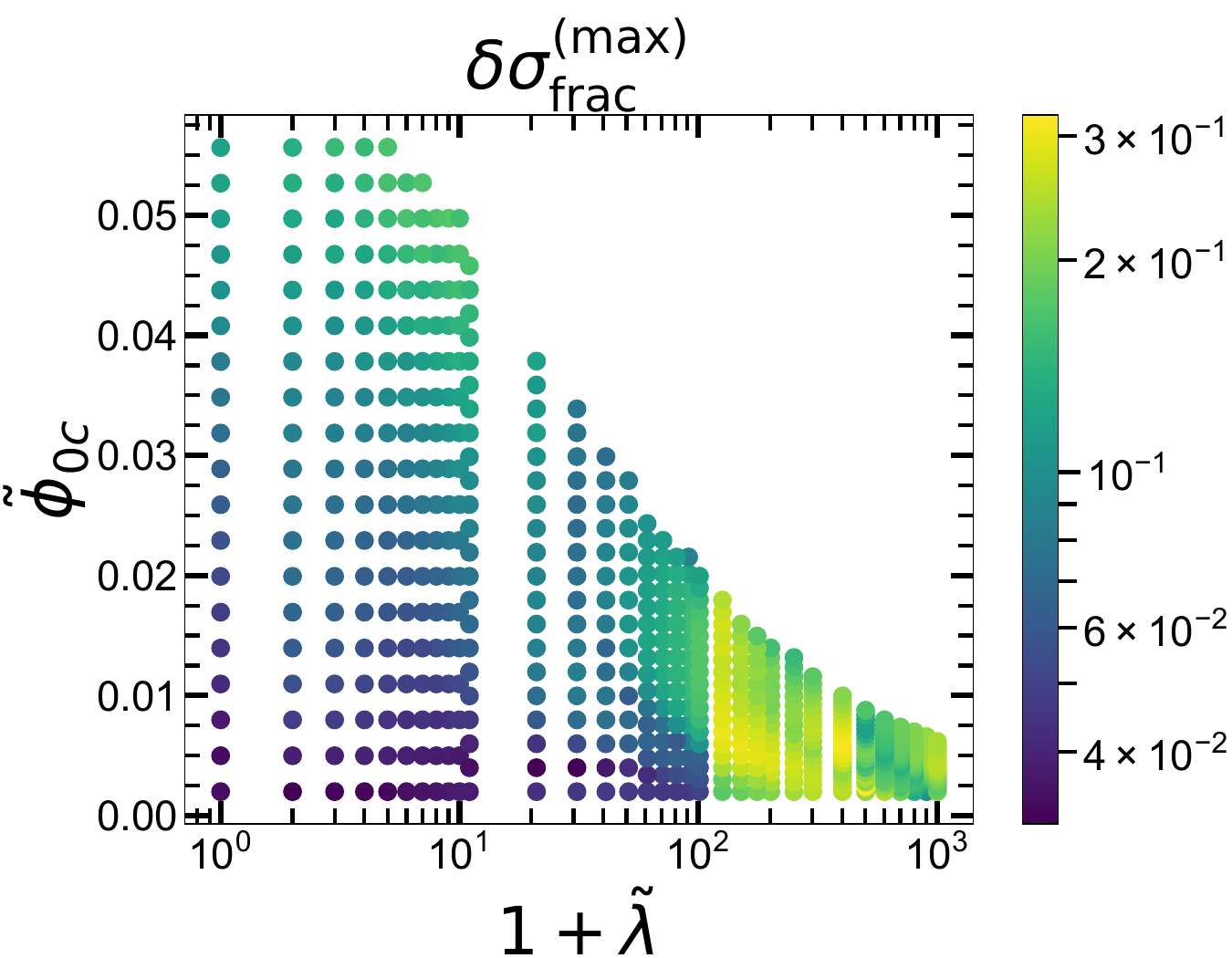}
         \includegraphics[width=0.32\textwidth]{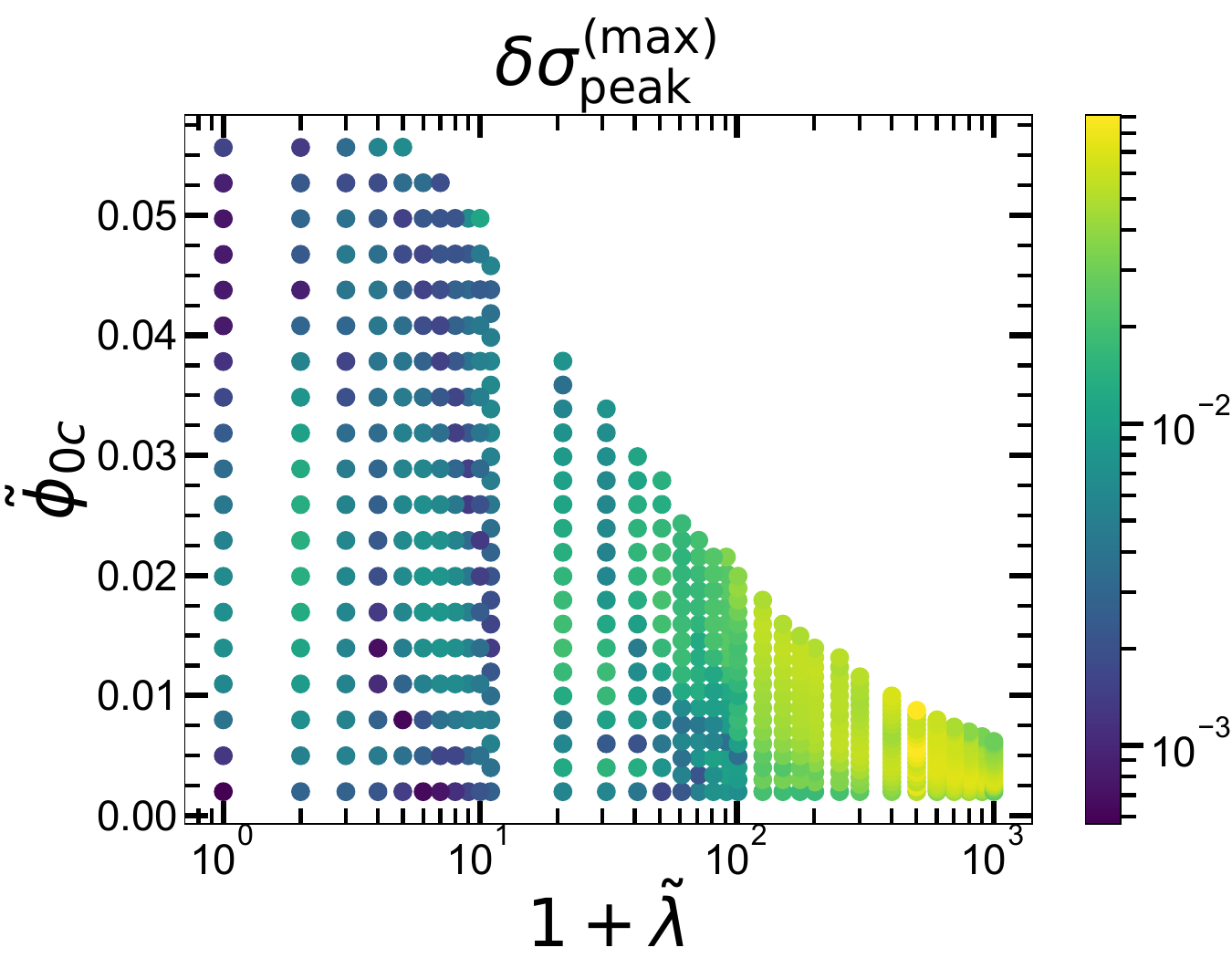}
         \includegraphics[width=0.32\textwidth]{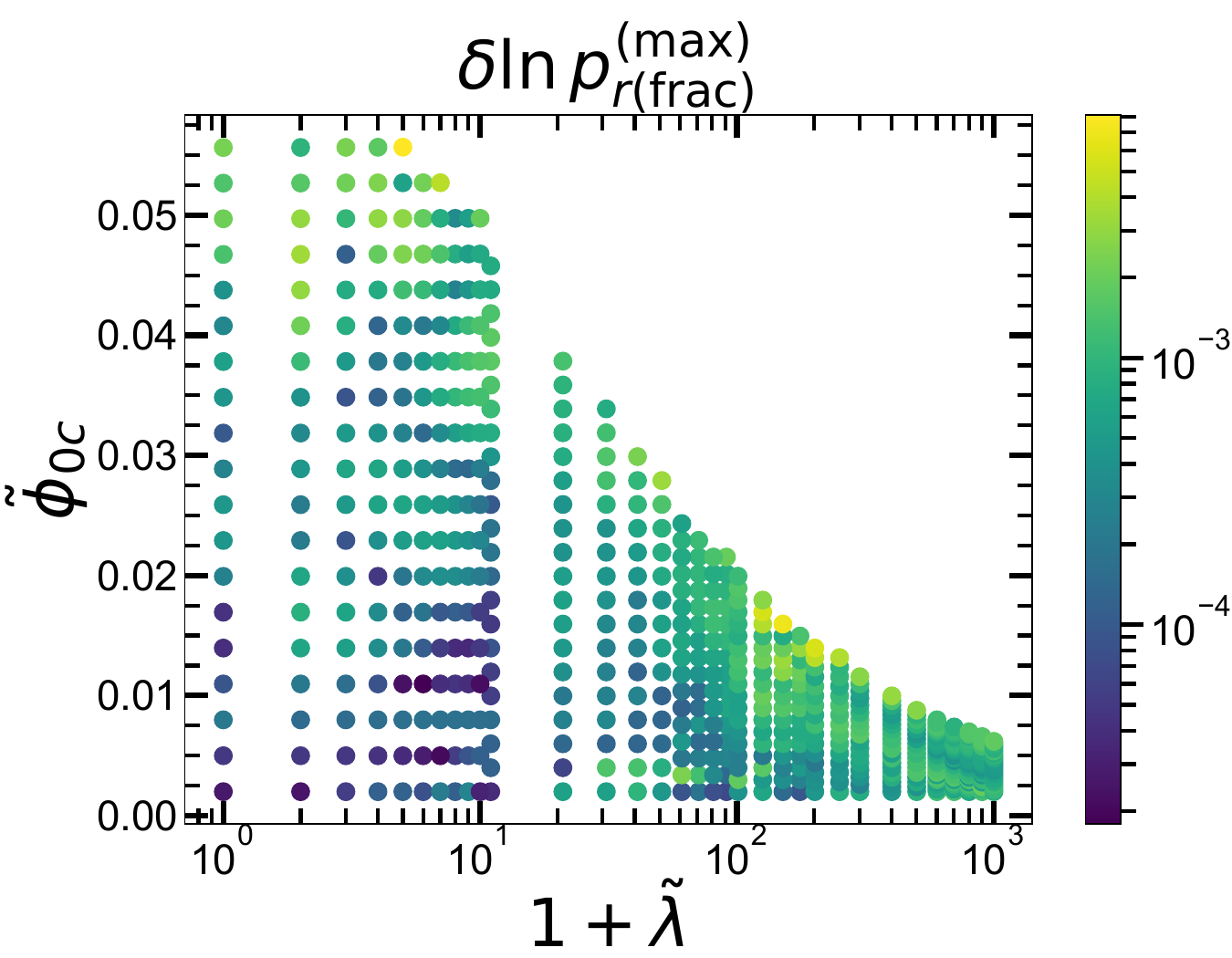}
        \caption{Maximum fractional errors of our fitting formulas. The left and center panels show the errors associated with the anisotropy fit given by Eq. \eqref{eq:sigma_fit}, while the right panel shows the errors for the equation-of-state fit given by Eq. \eqref{eq: EOSfitform}. The left and right panels display the maximum fractional error defined in Eq. \eqref{eq:frac_error}, whereas the center panel shows the peak normalized error defined in Eq. \eqref{eq:peak_error}. The vertical axis denotes the central scalar field value, $\tilde{\phi}_0$, expressed in Planck units. Note that the plots have different color scales.}
    \label{fig:errorplotscombined}
\end{figure*}

We next construct 2D fits for the above anisotropy constants in terms of $(\tilde \lambda, \tilde \phi_{0c})$ in the form
\begin{align}    \label{fitform} 
y = \sum_{i=0,1,2,4,6}\biggr[&a_{0i}+a_{1i}\tilde{\lambda}+a_{2i}\tilde{\lambda}^2+a_{3i}\tilde{\lambda}^3 \nonumber \\&+a_{4i}\tilde{\lambda}\ln(\tilde{\lambda}+1)  +\frac{a_{5i}}{\tilde{\lambda}+1} \biggl]\tilde \phi_{0c}^{i/2},
\end{align}
with $y=[\ln(-\beta_0), \ln(-\beta_1), \ln(\beta_2), n]$ (though there is no $\ln (\beta_2)$ in the range $100 \leq \tilde \lambda \leq 500$). These fits were obtained using the Levenberg-Marquardt algorithm~\cite{83b09f23-b20e-3617-8f72-24765b713f7b}. The fitting coefficients $a_{ij}$ for each regime in $\tilde \lambda$ are provided in a Mathematica notebook in~\cite{OnlineLink}. In the same repository, we also provide a ready-to-use TOV solver in Python that directly applies these fitting expressions in the construction of anisotropic fluid models for boson stars.

To quantify the accuracy of the fit, we compute two fractional error functions: 
\begin{align}
\label{eq:frac_error}
    \delta \sigma_\mathrm{frac}(\tilde r) =& \frac{|\sigma_\mathrm{num}(\tilde r) - \sigma_\mathrm{fit}(\tilde r)|}{\sigma_\mathrm{num}(\tilde r)}, \\
    \label{eq:peak_error}
    \delta \sigma_\mathrm{peak}(\tilde r) =& \frac{|\sigma_\mathrm{num}(\tilde r) - \sigma_\mathrm{fit}(\tilde r)|}{\sigma_\mathrm{peak}}.
\end{align}
The first one is the standard fractional error of the fitting form $\sigma_\mathrm{fit}(\tilde r)$ from the numerical data $\sigma_\mathrm{num}(\tilde r)$, while the second one is normalized by the peak value $\sigma_\mathrm{peak}$ of $\sigma_\mathrm{num}(\tilde r)$. The reason for introducing the latter is that the first fractional error typically reaches its maximum value near the center of the star, where $\sigma_\mathrm{num}(\tilde r)$ is small compared to its peak value. The left and middle panels of Fig.~\ref{fig:errorplotscombined} present the maximum values of the above two error estimates against $\tilde \phi_{0c}$ for selected  $\tilde\lambda$. Note that in Fig.~\ref{fig:errorplotscombined} we do not have a consistent uniform grid in the $\tilde \phi_{0c}-\tilde \lambda$ plane because we tried to use the minimum necessary number of grid points to construct accurate two-dimensional fits (Eq.~\eqref{fitform}).

\subsection{Equation of State}
\label{eos}

To complete the phenomenological fluid modeling of boson stars, we also need to provide a form of the equation of state, namely the radial pressure versus the energy density. For this work, we employ a quartic polynomial fit in log space given as 
\begin{equation}
\label{eq: EOSfitform}
   \ln p_r = \sum_{k=0}^4 q_k  \left(\ln \rho\right)^k,
\end{equation}
where we fit for each $q_i$ using Eq. \eqref{fitform} including $i = 3,5$. To improve our accuracy with these two parameter fits, we split our fitting regimes into $0 \leq \tilde{\lambda} \leq 100$ and $100 \leq \tilde{\lambda} \leq 1000$. The fitting coefficients $a_{ij}$ for each $q_i$ in each regime are provided in a Mathematica notebook in~\cite{OnlineLink}. 

\begin{figure*}
     \centering
             \includegraphics[width=0.4\textwidth]{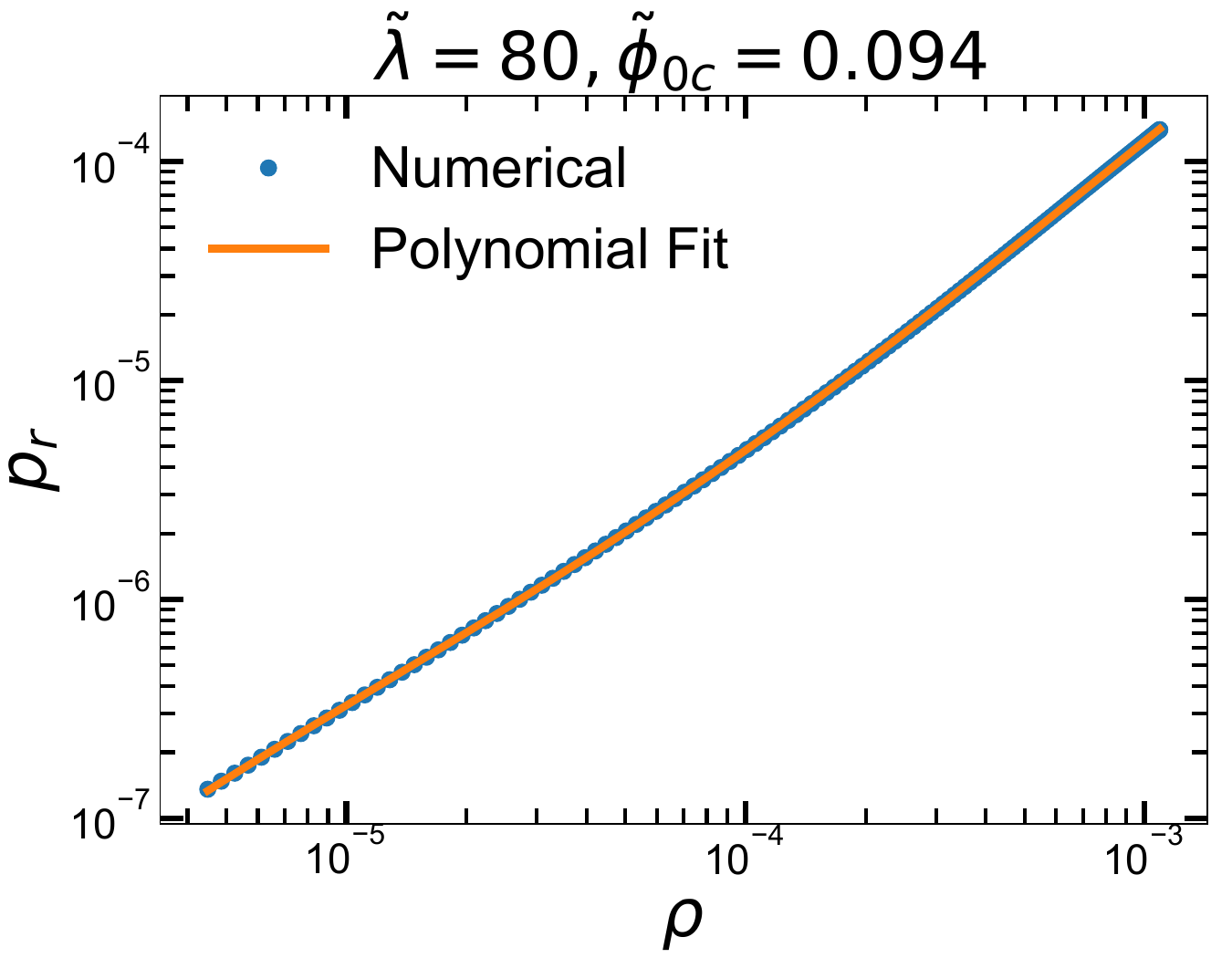}
         \includegraphics[width=0.4\textwidth]{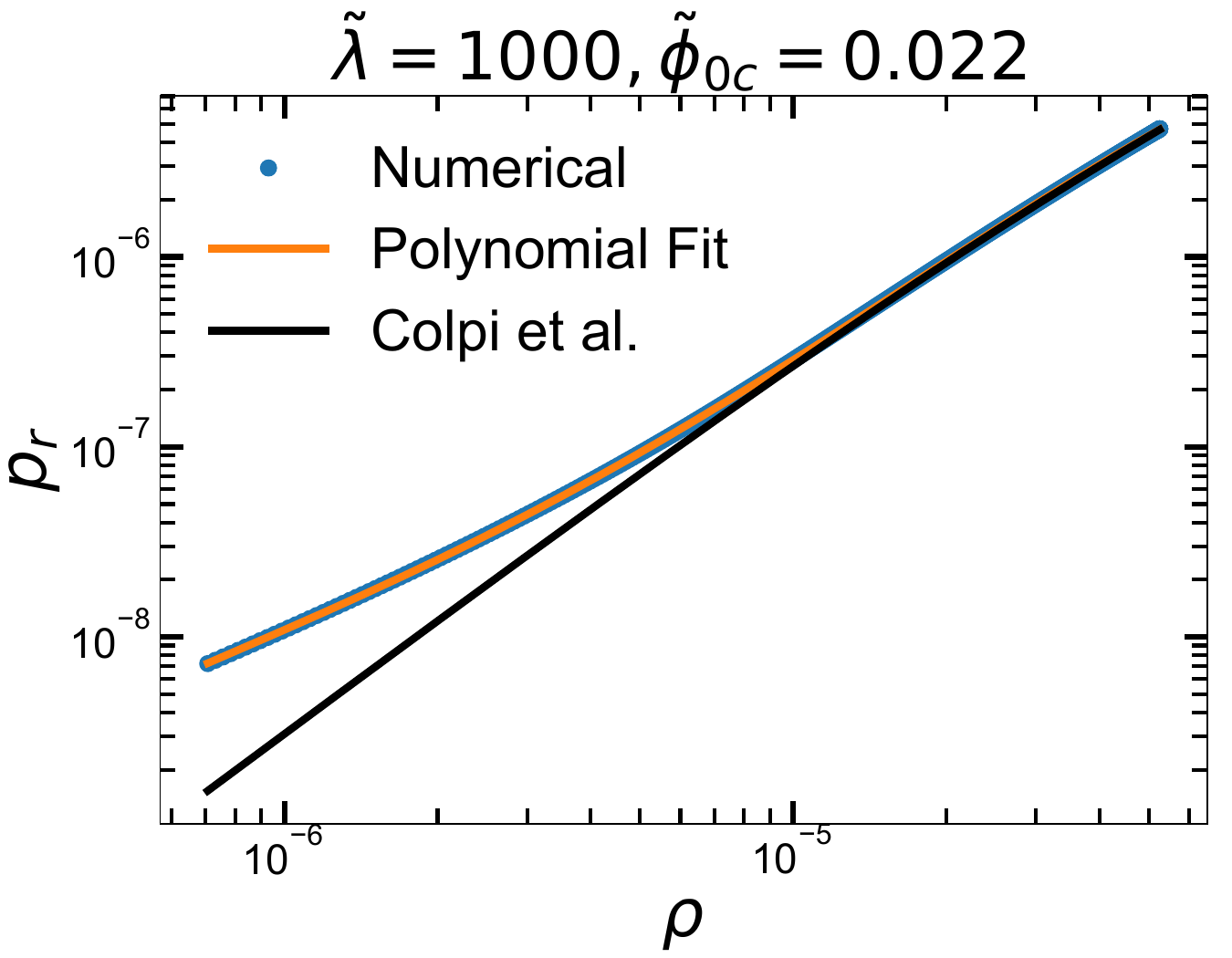}
        \caption{ 
        \emph{Left}: Effective equation of state for boson stars, together with a fit in Eq.~\eqref{eq: EOSfitform}.         
        \emph{Right}: Similar to the left panel but for a large value of $\tilde \lambda$. We also present 
        the equation of state formula found in \cite{PhysRevLett.57.2485}, derived in the $\tilde{\lambda} \rightarrow \infty$ limit.}
    \label{fig:EOS}
\end{figure*}

We demonstrate the accuracy of this fit in recovering our numerical results obtained using the relaxation method in Fig.~\ref{fig:EOS}.
The fractional errors in $\ln p_r$ for selected values of $\tilde{\lambda}$ are then shown in the right panel of Fig.~\ref{fig:errorplotscombined}.
In the right panel of Fig.~\ref{fig:EOS}, we compare our equation of state for $\tilde{\lambda} = 1000$ to the equation of state for high $\tilde{\lambda}$ derived in  \cite{PhysRevLett.57.2485}:
\begin{equation}
    \rho = \frac{4}{9}\rho_0\left[\left(1+\frac{3}{4}\frac{\rho}{\rho_0} \right)^{1/2}-1\right]^2,
\end{equation}
where $\rho_0 = \mu^4/4\lambda$. As expected, this equation works well near the center of the star but diverges from the true solution near the edges of the star where anisotropy becomes more significant.

\section{Applications}
\label{Applications}

In this section, we reconstruct boson stars through the anisotropic fluid model developed in the previous section and study its accuracy by comparing the stellar mass and radius with those from full calculations with the relaxation method.
For the reconstruction of boson stars with the phenomenological fluid model, we solve the TOV equations modified to account for anisotropy
(see e.g.~\cite{Doneva:2012rd}): 
\begin{align}
    \label{eq:dMdR}
\frac{d \bar \mu}{dr} &= 4 \pi r^2 \rho\,, \\
\frac{d v}{dr} &= 2\frac{4 \pi r^3 p_r + \bar \mu}{r(r-2\bar \mu)}\,, \\
\label{eq:dPdR}
\frac{dp_r}{dr} &= -\frac{(4\pi r^3 p_r + \bar \mu) (\rho + p_r)}{r(r-2\bar \mu)} - \frac{2\sigma}{r}\,.
\end{align}
Here
\begin{equation}
    e^u = \left( 1 - \frac{2\bar \mu}{r} \right)^{-1}\,.
\end{equation} 

We solve the above equations, supplemented by the equation of state $\rho = \rho(p_r)$ and the anisotropy model $\sigma$. 

In order to use the fits for the equation of state, we need to determine the minimum and maximum densities, $\rho_{\rm min}$ and $\rho_{\rm max}$ for which the ${p}_{r}$ vs. $\rho$ relation is valid, i.e. we need to determine $\rho_{\rm min}$ and $\rho_{\rm max}$, as functions of $\tilde\phi_{0c}$ and $\tilde{\lambda}$. To do so, we obtain the fits $\rho^{\rm fit}_{\rm min}(\tilde\phi_{0c},\tilde{\lambda})$ and $\rho^{\rm fit}_{\rm max}(\tilde\phi_{0c},\tilde{\lambda})$, using the expression in Eq.~\eqref{fitform} (including $i = 3,5$). We compute the fractional error (similar to Eq.~\eqref{eq:frac_error}) for these fits, and we obtain a maximum of $\sim0.08$ and $\sim0.002$ for $\rho_{\rm min}$ and $\rho_{\rm max}$, respectively. 

\begin{figure*}
    \centering
    \includegraphics[width=0.8\linewidth]{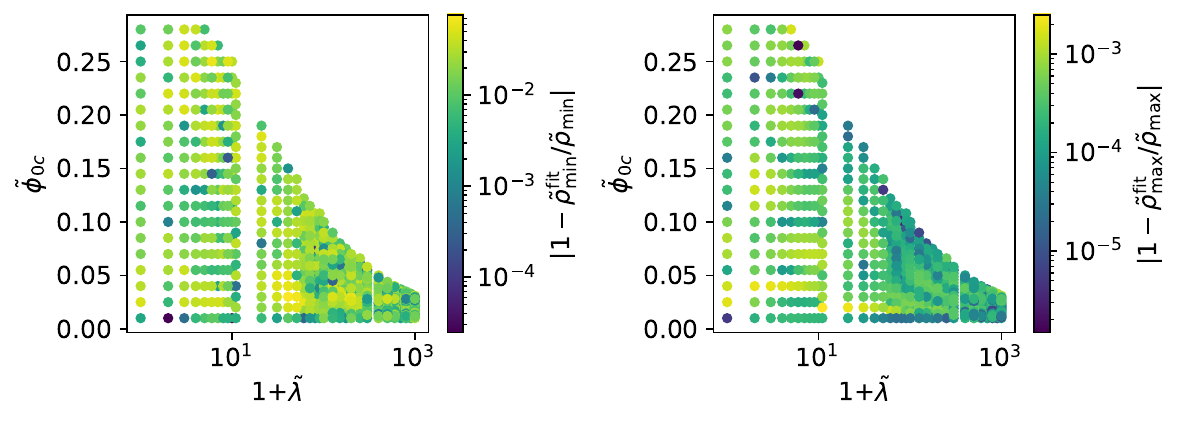}
    \caption{Maximum fractional errors of fits for $\rho_{\rm min}$ (left) and $\rho_{\rm max}$ (right) against $\tilde \lambda$ and $\tilde \phi_{0c}$.}
    \label{fig:minmaxrho}
\end{figure*}

We note that $\rho_{\rm max}$ and $\rho_{\rm min}$ can be treated as central and surfaces densities, $\rho_{c}$ and $\rho_{s}$.
This means that we determine the surface (and hence the radius) of a boson star from the phenomenological model as ${\rho}(R) = {\rho}^{\rm{fit}}_{\rm{min}}(\tilde{\phi}_{0c},\tilde{\lambda})$, which ensures we remain within the valid range of ${\rho}$ and ${p}_r$
used to create the fit for the equation of state. We have verified that the above two fits are valid for all values of $\tilde{\lambda}$ and $\tilde{\phi}_{0c}$ studied in this paper with the largest error of $\sim 7\%$ and $\sim 0.2\%$ for ${\rho}_{\min}$ and ${\rho}_{\max}$ respectively as shown in Fig.~\ref{fig:minmaxrho}.

Having our various fits in hand, we solve Eqs.~\eqref{eq:dMdR}--\eqref{eq:dPdR} under  appropriate boundary conditions near the stellar center $r=r_\epsilon$. Because the exponent $n$ in Eq.~\eqref{eq:sigma_fit} is larger than 1 in the parameter space studied in this paper,  the asymptotic behavior of $f(\tilde r)$ is given by $f(\tilde r_\epsilon) = \beta_0 \tilde r_\epsilon + \mathcal{O}(r_\epsilon^2)$ for $0 \leq \tilde \lambda \leq 500$ and $f(\tilde r_\epsilon) \approx \beta_0 \tilde r_\epsilon^{c_2}$ for $500 \leq \tilde \lambda \leq 1000$, with $c_2 = 1.05$. If we approximate it to $c_2 \approx 1$, the asymptotic behavior of $f(\tilde r_\epsilon) = \beta_0 \tilde r_\epsilon + \mathcal{O}(r_\epsilon^2)$ becomes valid for all  values of $\tilde \lambda$ studied in this paper. In this case, the asymptotic behavior of $\sigma_{0}$ and other quantities are given by
\begin{align}
    \rho (r_\epsilon) &= \rho_c + \rho_2 r_\epsilon^2 + \mathcal{O}(r_\epsilon^3)\,, \\
       \sigma =& \sigma_{0} \, r_\epsilon^2 + \mathcal{O}(r_\epsilon^3)\,, \\
       \label{eq:pr}
    p_r = & p_c - \left[\frac{2 \pi}{3} (3 p_c+ \rho_c) (p_c + \rho_c)+\sigma_{0}
\right] r_\epsilon^2+\mathcal{O}(r_\epsilon^3)\,, \\
\bar \mu(r_\epsilon) &= \frac{4\pi}{3} \rho_c r_\epsilon^3 + \mathcal{O}(r_\epsilon^5)\,, \\
v(r_\epsilon) &= v_c + \frac{4\pi}{3} (3 p_c+ \rho_c) r_\epsilon^2 + \mathcal{O}(r_\epsilon^4)\,, 
\end{align}
where $p_c$ is the central radial pressure while $\rho_c$ and $\rho_2$ are determined through the effective equation of state. Here $\sigma_{0}$ is given by
\begin{equation}
    \sigma_{0} = -\frac{2 \pi  \beta_0 (p_c+\rho_c) (3 p_c+\rho_c)}{3(2 \beta_0+1)}\,,
\end{equation}
and Eq.~\eqref{eq:pr} simplifies to
\begin{equation}
     p_r =  p_c +\frac{\beta_0+1}{\beta_0}\sigma_{0} r_\epsilon^2+\mathcal{O}(r_\epsilon^3)\,.
\end{equation}

\begin{figure*}
    \centering
    \includegraphics[width=\linewidth]{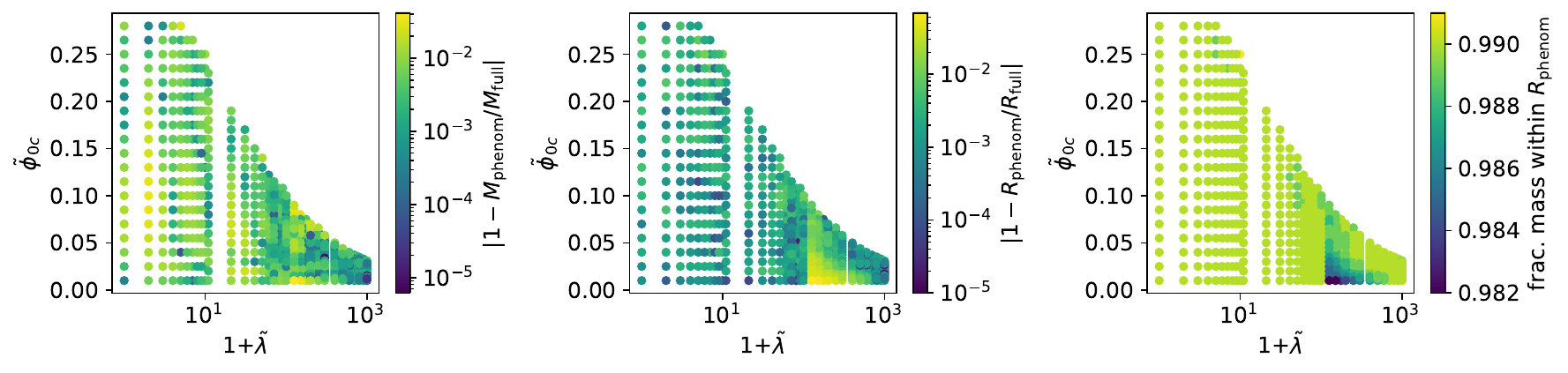}
    \caption{{\it Left}: Absolute fractional error between the mass of the boson star obtained using the phenomenological anisotropic fluid model, $M_{\rm phenom}$, and that obtained from a full calculation with the relaxation method, $M_{\rm full}$, as a function of $\tilde \phi_{0c}$ and $\tilde \lambda$. The model is able to reproduce the boson star mass from the full calculation with a maximum fractional error of $\sim 4\%$. {\it Middle}: Same as left, but for the absolute fractional error in the boson star radius, which can be reproduced with a maximum fractional error of $\sim 7\%$. {\it Right}: Same as left, but for the fractional mass enclosed, calculated from a full calculation with the relaxation method, within a sphere of radius $R_\mathrm{phenom}$, which lies between $\sim 0.98$ and $\sim 0.99$.}
    \label{mass_radius_frac}
\end{figure*}

Finally, the mass and radius for a boson star are determined as follows.
The stellar surface is determined as the location where the energy density reaches the surface value 
($\rho_{s}$ 
or equivalently $ \rho_{\min}$), 
according to $\rho^{\rm fit}_{\rm min}$ in Eq.~\eqref{fitform} (see Fig.~\ref{fig:minmaxrho}). However, besides using fits for $\rho_{\rm min}$, we also obtained fits for the radius from the full calculations, using Eq.~\eqref{fitform}. Thus, when solving the structure equations, we stop the integration at the radius inferred from the fit, if it corresponds to $\rho\geq\rho_{\rm min}$.
We then match the interior solution to the Schwarzschild spacetime due to Birkhoff's theorem. The stellar mass is determined by $M = \bar \mu(R)$.

In the left panel of Fig.~\ref{mass_radius_frac}, we present the results for the fractional error between the mass computed from our phenomenological fluid model $M_{\rm phenom}$  and the mass obtained from the full calculation with the relaxation method $M_{\rm full}$. We see that the former can reproduce the latter with a maximum fractional error of $\sim4\%$.

In the middle panel of Fig.~\ref{mass_radius_frac}, we show a similar fractional error for the boson star radius between the phenomenological model $R_\mathrm{phenom}$ and the full calculation $R_\mathrm{full}$. In this case, the maximum fractional error increases to $\sim 7 \%$. 

The radii of boson stars from the full calculations are defined as the value of $\tilde{r}$ such that 
$99\%$ of the mass is enclosed, although such a choice of the fractional mass is rather arbitrary. Let us next estimate this fractional enclosed mass from the full calculation with the relaxation method within a sphere of radius $ R _ {\ rm phenom} $ from the phenomenological model.  We show this result in the right panel of Fig.~\ref{mass_radius_frac}, where we see that the fraction varies between $\sim98\%$ to $\sim99\%$, and increases with ${\tilde{\phi}}_{0c}$ and $\tilde{\lambda}$. 

Figure~\ref{fig:MassRadius} compares the mass-radius relations of boson stars computed using the relaxation and phenomenological methods for selected values of $\tilde{\lambda}$. As shown, the phenomenological models are in excellent agreement with the results from the relaxation method.

\section{Conclusions}

In this paper, we constructed a ready-to-use phenomenological model for massive boson stars in the form of an anisotropic fluid. Such a model can be easily implemented to TOV solvers (such as the one we provided in~\cite{OnlineLink}) so that one can construct boson stars much more efficiently than using the computationally expensive relaxation method. We demonstrated the validity of the model by reconstructing spherically symmetric boson stars by solving the TOV equations modified to account for pressure anisotropy. We found that the stellar mass (radius) can be reproduced with a maximum error of $\sim 4$\% ($\sim 7$\%). The error on the radius is much larger because the definition of the stellar radius has some arbitrariness for boson stars as the scalar field extends all the way to spatial infinity. 

Several different avenues exist for future work.
First, one could improve the accuracy of the phenomenological model by introducing the compactified radial coordinate as done in~\cite{Sennett_2017} so that the boundary condition at $\tilde r \to \infty$ can be imposed more accurately. 
Second, it would be important to check the validity of the phenomenological fluid model for configurations beyond spherical symmetry.
One can study this by introducing various perturbations, such as static tidal perturbations for tidal deformability~\cite{Cardoso_2017,Sennett_2017} or dynamical perturbations for stellar oscillation frequencies~\cite{Flores_2019}. 
A framework already exists for studying static tidal perturbations~\cite{Yagi:2015hda} or stellar oscillations~\cite{Lau:2024oik} (see also~\cite{Doneva:2012rd,Mondal:2023wwo}) for anisotropic relativistic stars, though such a framework may need to be generalized to more accurately capture the effective pressure anisotropy for boson stars. Another interesting application to go beyond spherically symmetric configurations would be to introduce rotation, which is nontrivial as the spin angular momentum of rotating boson stars is known to only take discrete values~\cite{Kobayashi:1994qi}, and thus slowly-rotating solutions do not exist for boson stars (unless one performs an ``artificial'' averaging over the polar angle $\theta$~\cite{Delgado:2024fhc}). The application of the isotropic fluid model in the strong coupling limit to rotating boson stars has been discussed in~\cite{Vaglio:2022flq}.
Finally, it would also be useful to construct a similar phenomenological model for other types of boson stars, such as solitonic ones~\cite{Macedo:2013jja}.

\acknowledgments
E.M. is supported by NASA FINESST award 80NSSC25K0310 and a Montana Space Grant Consortium Graduate Fellowship. K.Y. acknowledges support from NSF Grant PHY-2309066 and PHY-2339969.

\appendix

\newpage

\bibliography{Main.bib}

\end{document}